# A fully automated urban PV parameterization framework for improved estimation of energy production profiles

B. Tian[*, 1], R.C.G.M. Loonen[1], R. Valckenborg[2], J.L.M. Hensen[1]

[1] Building Physics and Services, Eindhoven University of Technology, Postbus 513, 5600 MB Eindhoven, the Netherlands

[2] Netherlands Organization for Scientific Research (TNO), High Tech Campus 21, 5656 AE Eindhoven, the Netherlands

## Abstract

Accurate parameterization of rooftop photovoltaic (PV) installations is critical for effective grid management and strategic large-scale solar deployment. The lack of high-fidelity datasets for PV configuration parameters often compels practitioners to rely on coarse assumptions, undermining both the temporal and numerical accuracy of large-scale PV performance modeling. This study introduces a fully automated framework that innovatively integrates remote sensing data, semantic segmentation, polygon-vector refinement, tilt-azimuth estimation, and module layout inference to produce a richly attributed GIS dataset of distributed PV. Applied to Eindhoven (the Netherlands), the method achieves a correlation ($R^2$) of 0.92 with Distribution System Operator (DSO) records, while capacity estimates for 73% of neighborhoods demonstrate agreement within a $\pm 25\%$ margin of recorded data. Additionally, by accurately capturing actual system configuration parameters (e.g., tilt, azimuth, module layout) and seamlessly linking them to advanced performance models, the method yields more reliable PV energy generation forecasts within the distribution networks. Centering our experiments toward a high PV-penetration community, configuration-aware simulations help to reduce Mean Absolute Percentage Error (MAPE) of energy generation modeling by up to 160% compared to the conventional assumption-based approaches. Furthermore, owing to its modular design and reliance on readily available geospatial resources, the workflow can be extended across diverse regions, offering a scalable solution for robust urban solar integration.

## Keywords

---

[*] Corresponding author.
E-mail address: b.tian@tue.nl (B. Tian).

# Abbreviation list

| | |
|---|---|
| APE | Absolute percentage error [%] |
| CEC | California Energy Commission |
| CNN | Convolutional neural network |
| CPE | Cumulative percentage error [%] |
| CPW | Cumulative percentage width [%] |
| DSO | Distribution system operator |
| GCS | Geographic coordinate system |
| GIS | Geographic information system |
| GPB | Generation predictive band |
| GSD | Ground sample distance [m] |
| HD | Hausdorff distance [-] |
| IoU | Intersection over Union [-] |
| KNMI | Koninklijk Nederlands Meteorologisch Instituut |
| LiDAR | Light detection and ranging |
| MAE | Mean absolute error [$kW_p$] |
| MAPE | Mean absolute percentage error [%] |
| MAPW | Mean absolute percentage width [%] |
| MBR | Minimal-area bounding rectangle |
| MLP | Multilayer perceptron |
| PV | Photovoltaic |
| PVMM | PVMismatch |
| SAPM | Sandia array performance model |
| SCM | Semantic constraint module |

# 1. Introduction

Solar photovoltaic (PV) energy has emerged as an essential component in the global transition to low-carbon electricity generation [1]. Governments and industries worldwide continue to invest heavily in PV technologies due to their scalability, environmental benefits, and ever-decreasing costs [2]. In particular, PV installations in urban areas – such as rooftop systems on residential, commercial, and industrial buildings – are increasingly recognized for their potential to supply clean and decentralized power directly at the point of consumption [3,4]. These urban, decentralized PV systems provide not only environmental advantages but also help mitigate transmission losses by localizing power generation [5,6].

Despite the enormous potential, widespread integration of rooftop PV in urban settings also poses significant challenges for regional demand-supply balancing [7,8], and distribution grid planning and management of high PV penetration areas [9,10]. In response, a comprehensive and automated parameterization of distributed PV systems is crucial. This parameterization extends beyond simple detection, encompassing the detailed characterization of geographic locations, configuration specifications (e.g., module tilt, azimuth, and layout), and ultimately enabling an improved estimation of their energy production profiles. Such detailed information opens a promising solution to these challenges [11–13]. Additionally, robust PV system parameterizations further contribute to the decision-making process of auxiliary system (e.g., energy storage) investments [14,15] and energy pricing [16,17]. However, at urban scale, obtaining these PV system-specific parameters remains challenging [18,19], particularly due to a lack of widely accessible, open-source data resources [20,21]. Traditional efforts like manual surveys are time-consuming and difficult to scale [22,23], often only feasible for limited geographic regions [24]. In cases where accessible municipal data logs do exist, there are also possibilities the data is outdated, or part of the systems are mis-recorded [25,26].

Remote sensing, especially through satellite or aerial imagery, has emerged as a highly promising avenue for characterizing urban PV [27]. The rapid increase in spatial resolution and availability of satellite imagery, combined with advances in machine learning, provides an unprecedented opportunity to automate not just the initial detection but the entire parameterization of PV installations [28]. This automation is vital for keeping pace with the rapid growth of rooftop PV adoption in cities and ensuring that planning bodies have access to accurate and up-to-date data for applications like improved energy yield forecasting.

## 1.1 Urban PV parameterization: Beyond simple detection

In recent years, numerous studies have leveraged machine learning algorithms to identify the presence and location of solar arrays in high-resolution satellite or aerial images. These efforts have laid important groundwork for the initial step of PV system identification. Prior works often employed rule-based or heuristic image-processing techniques – such as edge detection [29,30], morphological erosion [31,32], and color thresholding [33,34] – to distinguish PV panels from the surrounding rooftops and grounds. In parallel, researchers also explored supervised statistical classifiers like support vector machines [35,36] and random forests [37–39] to learn characteristic features of RGB pixel variations from annotated datasets [40]. Although these approaches demonstrated the feasibility of automated PV detection, their generalization capability was frequently limited by factors such as complex rooftop shapes, PV material properties, and variations in lighting conditions [27,41].

With the rising popularity of deep learning, researchers began to explore convolutional neural networks (CNNs) [42–44] and other architectures for more robust image segmentation and classification tasks. For example, Malof et al. [45] combined a CNN with random forest classifiers to detect PV arrays in aerial imagery, achieving substantial improvements in accuracy compared to earlier rule-based methods. Kausika et al. [46] utilized TernausNet, a modified version of the U-Net architecture, to automate the mapping of PV installations in the Netherlands. To enhance the robustness of the detection results, two model variations were trained – one specifically designed for identifying modules on small buildings and the other for large buildings. Other projects have also leveraged prior semantic knowledge to enhance detection accuracy. For instance, DeepSolar [47] employed a transfer learning approach by fine-tuning a pre-trained Inception-V3 model on a large, PV-specific dataset, allowing the network to reuse general image features learned from diverse domains. Meanwhile, PV Identifier [48] integrated a Semantic Constraint Module (SCM) to incorporate rooftop context and solar module geometry, thereby refining the PV feature extraction and improving overall detection performance, showcasing the potential for developing high-resolution solar deployment databases. More recent developments have introduced stable diffusion inpainting techniques to produce synthetic rooftop PV samples [49], demonstrating how data augmentation approaches can address the scarcity of real-world training samples and enhance CNN-based segmentation robustness.

However, a significant limitation of most existing studies is their primary focus on merely identifying the spatial footprint of PV installations – i.e., confirming whether PV modules are present on a rooftop or not – without providing further installation parameters such as tilt, azimuth, array configuration, and the array's physical dimensions [27,37,50,51]. These studies, while valuable for initial PV detection, often stop there, lacking the subsequent analysis of more detailed system-specific parameters and a demonstration of how this detection contributes to solving real-world engineering problems. Those additional parameters are critical to robustly simulate real-time PV power generation [52], performance ratio [53], and seasonal production profiles [54]. For example, even small deviations in tilt or orientation can lead to notable changes in annual energy production [55]. Consequently, these "detection-only" studies do not construct an applicable PV parameterization pipeline and remain distant from practical applications.

The novelty of our work lies in the steps that follow the initial PV detection, specifically in developing a comprehensive, automated framework for urban PV parameterization. Few existing methods based on aerial imagery focus exclusively on azimuth estimation, typically by analyzing the rotation of segmented PV arrays [56,57] or deriving angles from rooftop ridges [58]. While such approaches succeed in identifying cardinal orientations, they offer no direct mechanism to infer installation tilt angles. A promising alternative for capturing both azimuth and tilt involves leveraging high-resolution 3D city models [59,60], which incorporate detailed geometric information on building roofs. However, publicly available 3D datasets remain scarce in many regions, and generating them on demand can be both labor-intensive and costly [61,62]. Another line of research derives the 3D orientation of PV systems using time-series measurements of electricity generation [20,63,64]. By examining the shape of daily production curves, it is possible to approximate tilt and azimuth angles. Yet, this strategy presupposes both the precise geographical (environmental) contexts of PV systems and widespread access to their generation data, which is generally unavailable for the majority of small- and medium-scale urban installations.

Additionally, existing satellite- or aerial- imagery-based capacity estimation methods face the difficulty of accurately scaling up from individual buildings to entire districts or metropolitan areas. Prior works have attempted to approximate installed capacity by correlating identified PV footprints with assumed module efficiency [25,65,66] or average watt-peak per unit area [67–69]. However, without access to additional installation details, these simplified approaches rarely account for array layout variations arising from partial rooftop obstructions or complex rooftop geometries [70]. As a result, estimation uncertainties can multiply rapidly when applying or upscaling such approaches to complex urban contexts [71,72]. Moreover, conventional PV information collection strategies often fragment the process into separate steps – such as PV footprint detection, capacity determination, and ultimate power estimation – without seamlessly integrating them into a unified pipeline. In practice, local authorities and utilities need both urban-scale coverage and site-specific accuracy to manage increasingly complex and decentralized electricity networks [73,74]. Achieving robust, citywide capacity assessments therefore demands a more comprehensive solution that couples automated PV detection with detailed system characterization and refined performance modeling – all within a single, scalable framework. Our work aims to provide such an integrated framework, demonstrating its application for improved PV energy profile modeling, a novelty that enhances the practical utility of PV parameterization datasets for dynamic generation forecasting.

## *1.2 Research objectives*

Recognizing these challenges and gaps, the present work introduces a novel, fully automated urban PV parameterization framework designed for improved estimation of energy production profiles. This framework encompasses all steps from initial detection to the collection and updating of system-specific data, as well as the final integration into advanced PV power models. Specifically, the primary objectives of this research are to:

- Develop multi-stage PV system-specific parameter estimation techniques and integrate them into a comprehensive, scalable framework that moves significantly beyond existing detection-only approaches.

- Demonstrate the practical utility and reliability of the generated urban PV parameterization dataset by validating the automatically derived parameters, particularly installed capacity, against ground-truth data from local energy utilities, thereby establishing its suitability for large-scale, real-world deployment.

- Illustrate the significant impact of this detailed parameterization by integrating the enriched PV dataset into advanced energy models to achieve an improved estimation and forecasting of urban-scale solar energy production profiles. This highlights the novel contribution of our framework in providing actionable data for enhanced dynamic energy management and grid planning strategies.

By achieving these objectives, the methodology offers an end-to-end solution for generating comprehensive PV system datasets on the city scale, which are directly applicable to enhancing energy modeling, forecasting, and informed decision-making in urban energy systems.

*1.3 Paper structure*

The remainder of this paper continues with a detailed description of the proposed parameterization workflow in Section 2, which explains the PV detection model, array configuration parameters extraction routines, and the capacity estimation procedure. Section 3 then presents the validation methods and metrics, drawing comparisons between the automated results and ground truth data. In Section 4, a demonstration study showcases the pipeline's real-world application for improved estimation of system energy generation profile within the selected urban regions, followed by a discussion of key significances, limitations, and outlooks in Section 5. Finally, Section 6 concludes the work and research outcomes.

## 2. Methodology

In the present study, remote sensing-based computer vision techniques are used to parameterize the adoption and specific installation details of rooftop PV power generation. The ultimate output is an enriched urban PV system dataset, ready for integration into advanced energy modeling. As illustrated by Figure 1, the parameterization workflow consists of five main steps: dataset preparation, PV array detection, installation tilt-azimuth angle calculation, array module layout prediction, and integration in PV power models for energy profile generation. Each step builds on the previous one to ensure a comprehensive and cohesive pipeline.

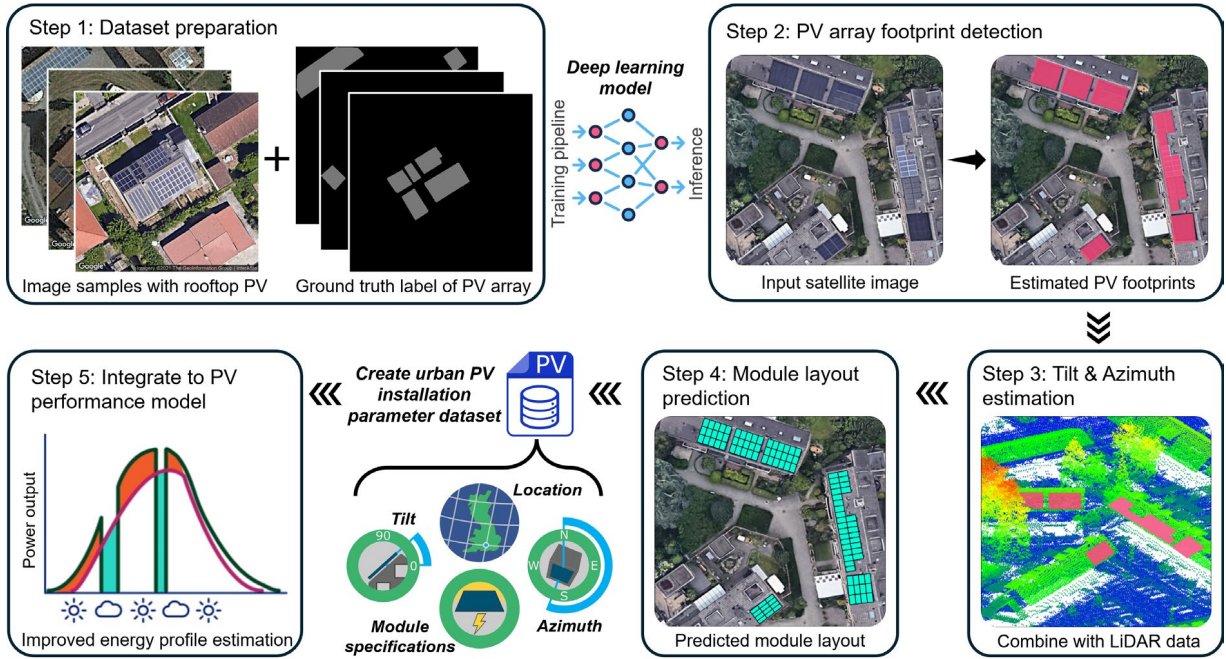

Figure 1. Diagram of the proposed urban PV parameterization framework.

## 2.1 Dataset preparation

In the proposed workflow, PV array detection model serves as the foundation for subsequent parameter estimations. However, development and validation of such a detection model fundamentally relies on a diverse and accurately labeled dataset. Therefore, as the initial step in our methodology, we aggregated labeled satellite imagery from multiple countries worldwide [75–77]. This diverse dataset contributes to enhance the generalizability of the entire framework across various urban scenes, e.g., building types, lighting conditions, and backgrounds. Table 1 illustrates the final composition of the dataset, which comprises images of different ground sample distances (GSD): 0.1 m, 0.3 m, and 0.8 m. The GSD reflects the spatial resolution of the imagery, and having a range of resolutions allows the model to learn robust features across different data qualities and environmental settings.

Tabel 1. Statistic of aggregated open-sourced dataset.

| GSD [m] | Sample source [-] | Sample number [-] | Proportion [%] | Ref. |
| --- | --- | --- | --- | --- |
| 0.1 | China | 645 | 2.5 | [75] |
| 0.1 | Denmark | 880 | 3.5 | [76] |
| 0.1 | France | 13303 | 52.0 | [77] |
| 0.3 | China | 2308 | 9.0 | [75] |
| 0.3 | France | 7685 | 30.0 | [77] |
| 0.8 | China | 763 | 3.0 | [75] |

In all cases, data samples were provided as square image tiles accompanied by pixel-level annotations for PV arrays. For the higher-resolution datasets at 0.1 m and 0.3 m GSD, which provide rich spatial detail, image tiles were 256×256 pixels; this size is common for deep learning model inputs and balances the detailed information content with computational manageability. In contrast, for the lower-resolution 0.8 m GSD images, where individual PV arrays appear smaller in pixel dimensions, larger tiles of 1024×1024 pixels were used. This approach for the coarser GSD ensures that each tile still encompasses a sufficient geographic extent to capture adequate contextual information and a variety of rooftop configurations, which is crucial for effective model training on features that are less distinctly resolved.

Before splitting the dataset, the entire pool of labeled tiles was randomly shuffled to minimize any bias associated with spatial ordering. Subsequently, the data was divided into training, validation, and test subsets in a 60 %-20 %-20 % ratio. This approach guaranteed a sufficient volume of labeled images for model learning (training set), hyperparameter tuning and performance monitoring (validation set), and final accuracy assessment (test set).

By combining multiple regional sources, leveraging diverse GSD values, and structuring the data rigorously, we established a comprehensive, high-variance dataset. Such diversity is instrumental in enabling the PV array detection model to adapt well to new urban environments and varying image qualities in real-world applications.

*2.2 PV array detection*

The accurate detection of PV arrays from satellite imagery forms a critical step in our overall workflow. By identifying the exact footprint of PV installations on building rooftops or open spaces, we establish the foundation for subsequent geometric and systematic data analyses. The detection of PV arrays is composed of two steps: (1) generating pixel-level masks through semantic segmentation, for which we adapt an existing high-performance model, and (2) our novel method for refining those masks into vector polygons for more practical representation and analysis.

### 2.2.1 PV mask segmentation

To detect the widespread PV installations, binary segmentation masks that indicate whether each pixel belongs to a PV array or not should be first generated across the input imagery. As described in Section 1.1, since PV binary segmentation is already a well-established field, we adopt the well-performing SegFormer model to handle this initial detection task. SegFormer is a transformer-based semantic segmentation model, which have been reported to achieve state-of-the-art performance on various benchmark datasets [78,79]. Briefly, SegFormer's encoder integrates multi-stage transformer blocks that capture long-range dependencies and local details simultaneously. An all-MLP decoder then aggregates feature maps from different encoder stages, producing pixel-wise predictions for each location in the image. As shown in Figure 2, in our workflow, the output of this model represents a binary mask that classifies each pixel as "PV array" or "background." Readers can refer to the original SegFormer publication by Xie et al. [78] for additional technical details.

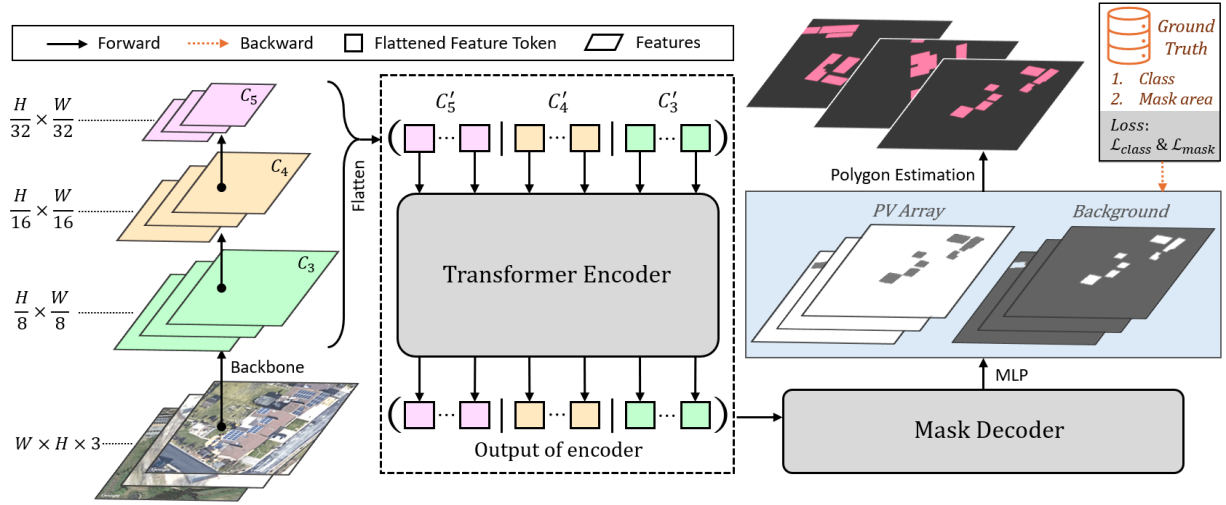

Figure 2. Diagram of PV detection stages, wherein the input image tiles are passed through the SegFormer backbone, generating hierarchical feature maps which are transformed into tokens for the transformer encoder. After the all-MLP decoder combines the multi-level feature information, the output mask is derived.

In our experiments, we initialize the SegFormer encoder with ImageNet-pretrained weights to leverage the general image features learned from large-scale datasets. During training, we use the Adam optimizer with a weight decay of 0.01 and set the initial learning rate to $5\times10^{-4}$. The network is trained for 2,000 iterations on a single NVIDIA RTX 3090 GPU, which is sufficient to converge on our labeled satellite imagery dataset (see Section 2.1 for dataset preparation).

Additionally, we employed focal loss to address the class imbalance typical in overhead imagery, where PV panels represent relatively few pixels compared to background. Focal loss effectively reweighs hard-to-classify examples [80], ensuring that the model concentrates on refining boundary regions of PV arrays. In practice, this helps minimize the mismatch between predicted masks and ground-truth annotations, thereby reducing the risk of false positives or missed detections.

The output segmentation masks provide the foundation for subsequent steps, where we convert pixel-level predictions into higher-level representations such as polygons (Section 2.2.2) and ultimately derive more detailed system parameters.

### 2.2.2 PV polygon estimation

Although the segmentation masks produced by SegFormer provide a clear visualization of where PV arrays are located in an image tile, these pixel-based predictions do not directly encode geometric boundaries in a manner readily usable by downstream analyses or geographic information system (GIS) workflows. Besides, traditional pixel-based contouring methods often fall short in capturing the true, often rectilinear, nature of PV arrays or handling segmentation imperfections effectively. Therefore, we further introduce an innovative iterative refinement process to convert the binary segmentation masks into precise vector polygons with explicit geometric coordinates. This vectorization method is essential for bridging the gap between raw pixel-level detections and the high-fidelity geometric inputs required for subsequent unique parameter estimation techniques (such as tilt, azimuth, and module layout prediction) and spatial queries.

Figure 3 illustrates the developed approach for converting raw PV pixel masks ($M$) into polygon vectors ($P'$). In the first iteration, we isolate each connected group of PV pixels in the mask, then approximate the shape of each group with minimal-area bounding rectangles (MBR) $P_1$. Next, for the observed mismatched region between $P_1$ and $M$, we calculate the second group of MBRs (annotated as $D_1$ in Figure 3). Subsequently, the initial PV polygons $P'$ are obtained by subtracting $D_1$ from $P_1$. However, due to the complexity of actual array shapes, the mismatched areas $D_1$ sometimes contain further subregions of true PV pixels; in this case, more iterations of this MBR-based subtraction and addition are executed. As demonstrated by sample B and C in Figure 3, in each iteration, we either remove or retain polygons within mismatch areas, depending on whether they correspond to genuine PV footprints. Finally, the PV polygon $P'$ is refined by uniting all the valid partial polygons ($P_i \backslash D_i$) from multiple iterations into a single, topologically consistent polygon layer.

This iterative refinement operation, specifically designed to handle complex array geometries and segmentation imperfections by intelligently adding or subtracting regions based on MBR analysis, is a key contribution that ensures the generation of high-quality footprints. These accurate footprints are indispensable for our advanced parameter estimation modules. Additionally, polygon-level estimation results are georeferenced according to the known geographic coordinate system (GCS) of the satellite image, ensuring that each PV array can be mapped to real-world coordinates.

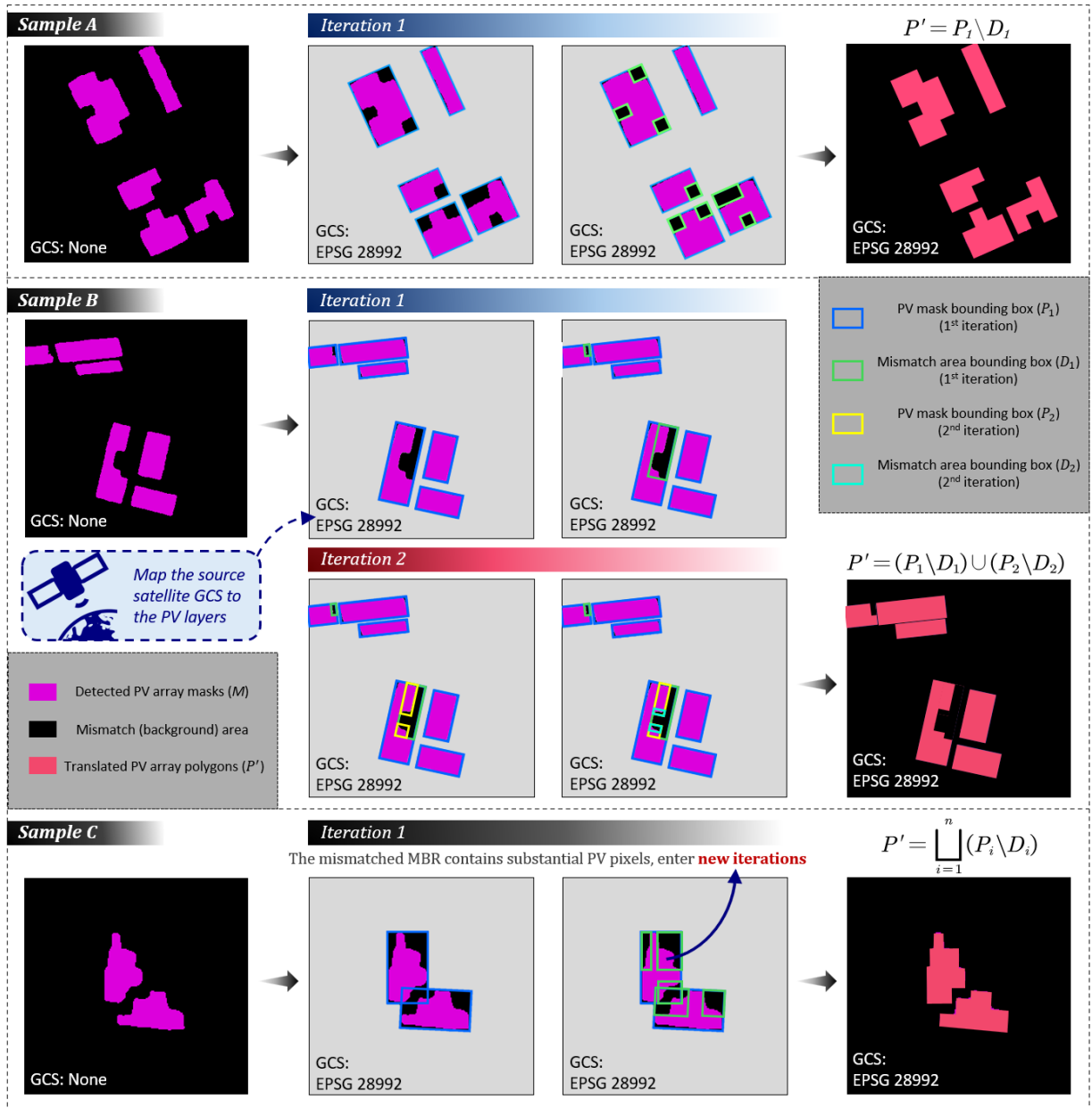

Figure 3. Diagram of PV array polygon estimation approach.

Through this novel iterative refinement process, the irregular and noisy boundaries inherent to segmentation masks are converted into a set of polygon vectors that accurately reflect each PV array's footprint. The final result is a collection of polygons that can be mapped, measured, and analyzed seamlessly in subsequent stages of our parameterization workflow or integrated into external GIS platforms.

### 2.2.3 Model performance

Evaluating the performance of our PV array detection pipeline involves examining both the raw mask predictions generated by SegFormer (Section 2.2.1) and the final polygon outputs derived from the iterative bounding-box decomposition (Section 2.2.2). In the present work, we adopt two well-established metrics: Intersection over Union (IoU) and the Dice coefficient, as they effectively capture how closely the predicted region aligns with the ground truth in both shape and location.

***Intersection over Union (IoU).*** As shown in Equation 1, given *A* is the set of predicted pixels (or predicted polygon area) and *B* is the set of ground-truth pixels (or reference polygon area). IoU indicates the proportion of overlap between the prediction and ground truth relative to their union. Higher IoU values imply more precise agreement in shape and location.

$$IoU(A,B) = \frac{|A \cap B|}{|A \cup B|} \qquad (1)$$

***Dice Coefficient.*** Often interpreted as the F1 score for segmentation (Equation 2), Dice emphasizes overlapping regions in a manner similar to IoU but can be more sensitive when the predicted and ground-truth areas differ substantially in size.

$$Dice(A,B) = \frac{2|A \cap B|}{|A| + |B|} \qquad (2)$$

Table 2 reports the IoU and Dice scores for both the raw PV mask segmentation and the refined array vector polygons estimation across testing imagery samples of varying GSD. Overall, the SegFormer-based segmentation demonstrates high accuracy in identifying PV installations, with IoU and Dice values remaining consistently strong at multiple spatial resolutions. While the 0.1m GSD data exhibit particularly high pixel-level accuracy – attributable to their clear depiction of rooftop details – coarser images at 0.8m GSD can still yield surprisingly robust mask metrics, possibly due to fewer spurious edges or small occlusions that could otherwise degrade segmentation quality.

Table 2. Performance scores of the PV array detection model.

| GSD [m] | PV array mask segmentation | | PV array polygon estimation | |
|---|---|---|---|---|
| | IoU [-] | Dice [-] | IoU [-] | Dice [-] |
| 0.1 | 0.93 | 0.96 | 0.81 | 0.89 |
| 0.3 | 0.88 | 0.93 | 0.76 | 0.86 |
| 0.8 | 0.91 | 0.95 | 0.92 | 0.96 |

Performance differences become more pronounced when transforming the masks into polygons. At fine resolutions (e.g., 0.1 m GSD), small boundary irregularities and complex rooftop geometries can slightly reduce the IoU for polygon outputs relative to mask-level predictions. In contrast, polygon-level performance at 0.8 m GSD remains notably strong, reflecting how simpler roof features in lower-resolution data can lead to more stable bounding-box refinements and fewer mismatch subregions.

These observations highlight a fundamental trade-off between capturing fine rooftop details on one hand and maintaining simpler, more rectilinear shapes on the other. Despite these nuances, results in Table 2 demonstrate that the proposed pipeline – combining advanced segmentation with iterative bounding-box decomposition – consistently produces high-quality PV detection layers regardless of resolution. This robustness is especially valuable for real-world deployment, where imagery often varies in quality depending on regional data availability.

## 2.3 Installation tilt-azimuth angle calculation

Although collecting the orthogonal PV footprints on the horizontal plane is essential, advanced modeling of expected energy yield further requires knowledge of the array's tilt and azimuth. Another novel and critical component of our automated parameterization framework is the methodology to derive these angles. Following the accurate PV footprint vectorization (Section 2.2.2), our approach leverages openly available urban point clouds, which are often more accessible than full 3D city models.

As illustrated in Figure 4, our approach derive these angles by overlaying detected PV polygon layers with openly available urban point clouds. Because both data sources share the same map projection, each polygon can be used to "clip" the three-dimensional points corresponding to that specific rooftop area. The extracted set of $(x, y, z)$ coordinates, here referred to as the "array-level pointset", thereby capturing the true geometry of the PV installation. This direct extraction of 3D geometry from point cloud data, guided by the precise 2D footprints generated by our novel vectorization method, is an advancement over methods relying solely on 2D imagery or requiring complex 3D model reconstruction. By finding the best-fitting plane and normal vector of individual array pointsets, we obtain a physically meaningful estimate of its tilt ($\theta$) and azimuth ($\gamma$). This capability to estimate both tilt and azimuth angles from readily available data sources significantly enhances the detail of the PV parameterization, which is often missing in other large-scale detection efforts.

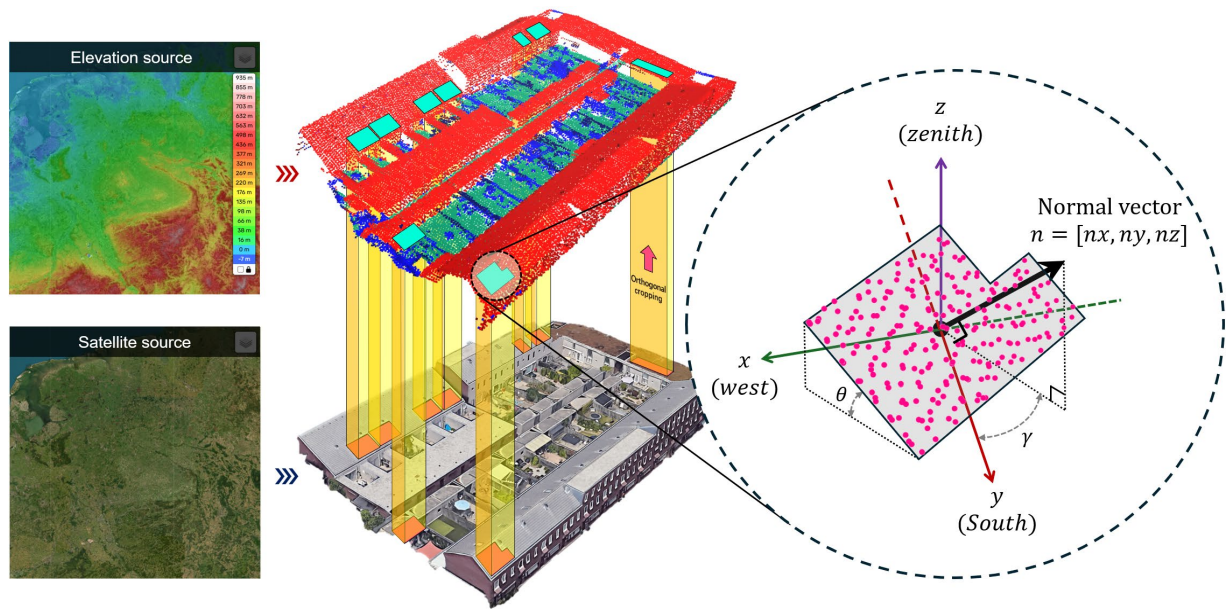

Figure 4. Diagram of calculating PV installation tilt-azimuth angle from LiDAR point cloud data.

## *2.4 Module layout prediction*

Once the tilt and azimuth are determined, our framework incorporates another novel stage to further refine the PV system characterization: the prediction of individual module layouts within each rooftop PV footprint. Moving significantly beyond simple area-based capacity estimates found in many existing approaches, this stage reconstructs a realistic module layout that unlocks further applications – including highly detailed capacity assessments based on actual module counts and configurations – as well as refined estimates of array geometry for advanced shading and performance modeling.

A key challenge in predicting module layout is that even though PV module sizes tend to be standardized, there is still considerable variation in module dimensions across different manufacturers and product lines. This variability makes correctly configuring modules into the target PV footprints and accurately determining the module count a non-trivial task. To address this diversity, our methodology involves fitting various standard module dimensions to evaluate different possible layout variations. Figure 5 illustrates the core pipeline: starting with the collected array polygons and their estimated tilt angles, a set of California Energy Commission (CEC) approved standard module templates [†] (listed in Appendix A) representing this common dimensional variability, is iteratively placed onto a "virtual grid" aligned to the polygon's minimal-area bounding rectangle (MBR). Each module template is tested in both portrait and landscape mounting orientations, thereby generating multiple candidate layouts per array.

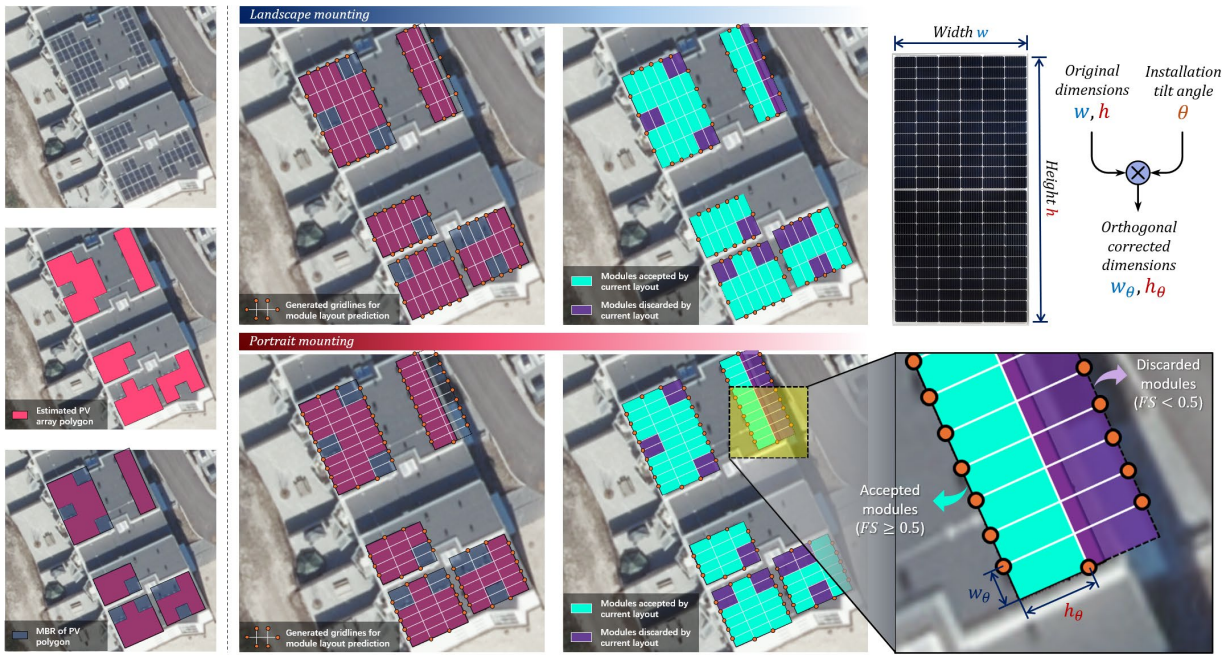

---

[†] https://solarequipment.energy.ca.gov/Home/PVModuleList

Figure 5. Predicting PV module layout over the collected array polygons, modules subject to higher fitness scores are accepted and preserved in the configuration grid.

The virtual grid approach accounts for the PV polygon's actual tilt $\theta$ by adjusting the module's height (or width) using $\cos(\theta)$. Grid cells partially or wholly lying outside the polygon are discarded based on a coverage ratio that checks whether most of a cell overlaps with the PV footprint. Accepted cells are merged into a multi-module polygon, producing one layout hypothesis. As our model evaluates all module types in both orientations, each PV array can yield up to 46 different layouts (23 module types ×2 mounting orientations).

To identify the most realistic solution, we define a matching score balancing area overlap and boundary alignment. Let $A$ be the candidate layout polygon and $B$ the ground-truth PV array polygon. We compute the IoU score to weigh their area overlap, together with the Hausdorff distance (HD) measuring the maximum boundary-to-boundary discrepancy. Higher IoU indicates strong coverage, whereas lower HD reflects closer geometric conformance. We then combine these into a single matching score $S$:

$$S(A,B) = \frac{IoU(A,B)}{1 + HD(A,B)} \qquad (3)$$

Conceptually, this score rewards shapes that match not only in coverage but also in boundary alignment. Figures 6 and 7 illustrate how arrays with high matching scores ($\geq 0.7$) fit the polygon almost perfectly, covering the actual system area without large gaps or overextensions. In contrast, low scores ($\leq 0.4$) typically indicate that the predicted layout either undershoots or overshoots the actual boundary, leaving unfilled areas or protruding well beyond the roof edge. By selecting the layout with the highest matching score, we ensure that the final configuration is both geometrically consistent and closely aligns with real-world PV arrangement practices. This detailed, module-specific layout prediction is a distinctive feature of our parameterization framework, offering a granular level of detail previously unattainable at large scales with automated methods.

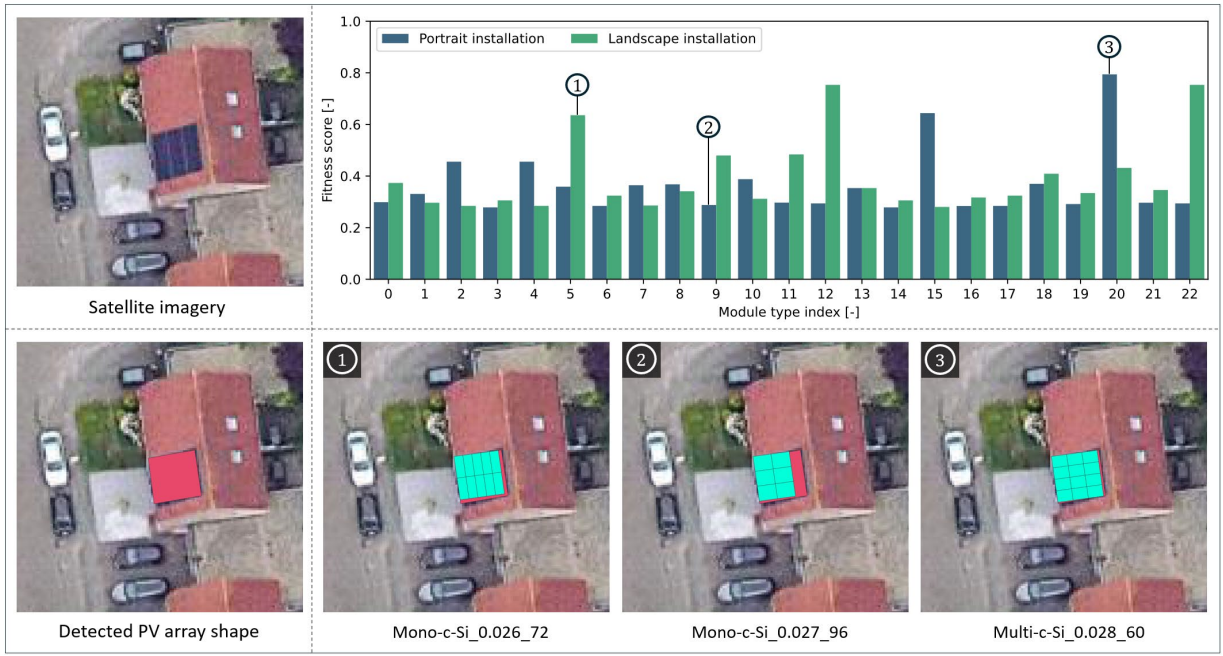

Figure 6. Example module layout hypothesis with different CEC standard module templates, the associated matching scores varied significantly.

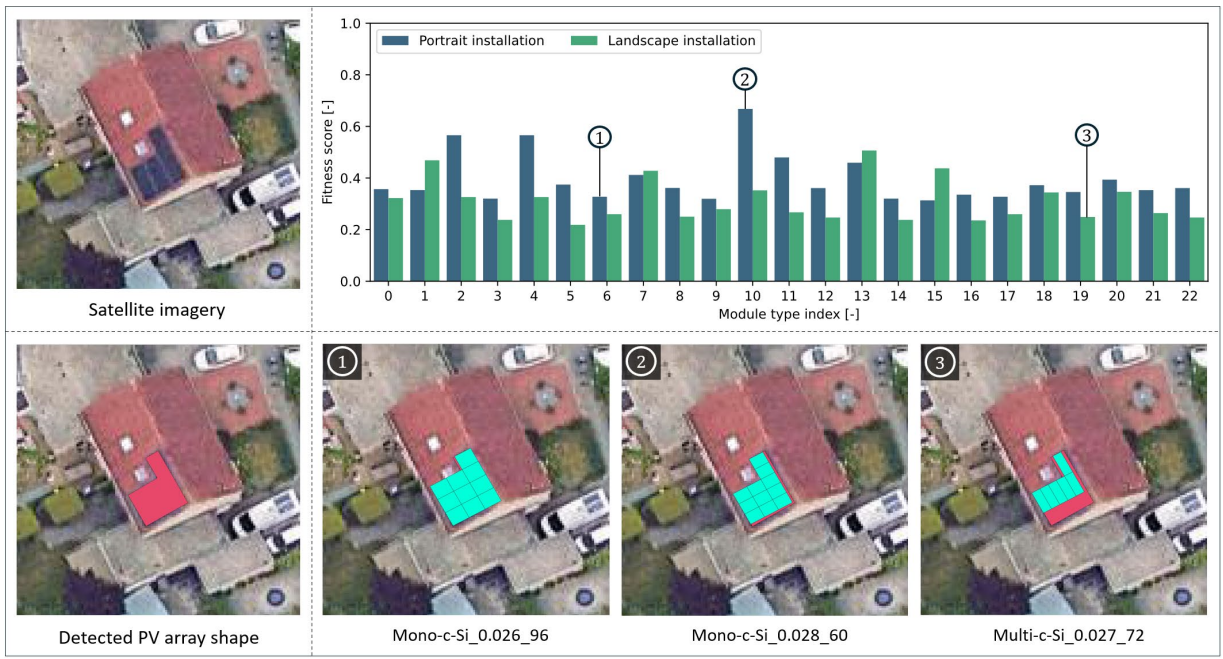

Figure 7. Example module layout hypothesis with different CEC standard module templates, the associated matching scores varied significantly.

After the best layout is chosen, the rated power of each placed module is summed to obtain the overall installation capacity. This yields an updated set of PV polygons in which each feature not only has tilt, azimuth, and geometric boundaries, but also a precise module-level arrangement and total installed power rating. In turn, this exceptionally enriched data layer, a direct output of our novel parameterization steps, affords extensive possibilities for energy yield forecasting and distribution network modeling.

# 3. Validation

To assess the reliability and practical relevance of the proposed parameterization framework, we conducted a validation exercise. This involved comparing our framework's urban-scale PV installation capacity estimates with officially recorded PV capacity data from the local Distribution System Operator (DSO). It is important to recognize at the outset that while such DSO data provides the best available large-scale reference, these records can themselves be subject to inherent uncertainties, including mis-registrations or outdated entries. Therefore, this validation aims not only to quantify the agreement between our estimates and the recorded data but also to interpret these comparisons within the context of these potential real-world data discrepancies. The objective is thus to evaluate if our framework provides robust and plausible estimations, rather than seeking a strict fitting to potentially imperfect records – framing this as a soft optimization challenge.

## *3.1 Experiment setups*

In this validation exercise, we focus on Eindhoven, a medium-sized city where decentralized rooftop PV has been rapidly expanding in the past decade. To obtain the necessary image inputs for the model, we retrieved satellite imagery from Google Maps API at zoom level 20, corresponding to an approximate GSD of 0.3 m. These images, captured in 2022, were systematically tiled into 256 × 256 pixel patches covering the entire urban area. Leveraging our comprehensive parameterization pipeline (Section 2), we identified all visible rooftop PV arrays within each tile and translated them into a unified urban PV GIS layer with detailed tilt, azimuth, module layout information, and installation capacity. Figure 8 illustrates these inference results at the city level, visually confirming that the distributed PV areas are well-identified across Eindhoven's municipal region, while the proposed layout estimation yields a realistic grid of modules for each array footprint.

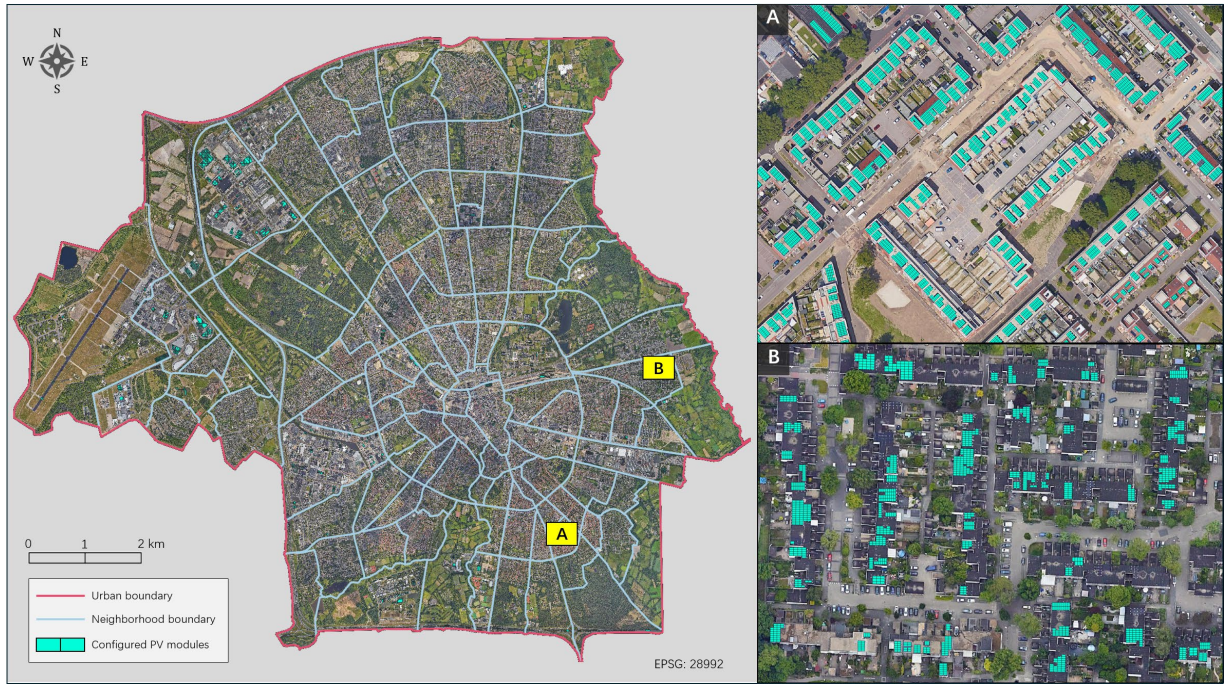

Figure 8. Model inference results of city Eindhoven, the Netherlands. Two municipal subregions are zoomed out to assist visualizations.

Urban-scale data recordings of installed PV capacity, which are updated semiannually and broken down by municipal subregions or neighborhoods [‡], were sourced from ENEXIS, the principal Distribution System Operator (DSO) in the region. While this represents valuable ground-truth information, it is acknowledged in system measurement studies that such administrative datasets from DSOs can sometimes exhibit discrepancies due to factors like registration lags, administrative errors, or unrecorded installations [26]. Because the satellite images used for our detection efforts were also taken in 2022, we retrieved the second-half 2022 data from ENEXIS to align with the imagery capture period. The dataset distinguishes between small-scale (residential and smaller commercial) and large-scale (heavy industrial or utility-scale) PV systems. While the small-scale entries are subject to a formal registration process, the large-scale data merely reflect regional inverter capacities, lacking sufficient fidelity for precise spatial comparisons. Consequently, only the small-scale connections are included in this validation. By aggregating our model's estimated capacities at the same neighborhood boundaries, we aim to compare the two sources in a consistent, area-based manner.

*3.2 Results*

To evaluate how closely our model-based estimates align with the recorded DSO neighborhood capacity data, we present two scatter plots (Figures 9 and 10) comparing observed (x-axis) versus predicted (y-axis) capacities across Eindhoven's various neighborhoods. In both figures, the diagonal line denotes perfect agreement between the model and the recorded data; deviations from this line will be analyzed considering potential error sources from both our estimation framework and the reference DSO dataset itself. In addition, density plots of installed capacities are shown on the top (secondary x-axis) and right (secondary y-axis) margins to characterize the distribution of neighborhood capacities in the recorded dataset and our predictions, respectively. Moreover, three quartile markers are highlighted along with the capacity distributions in both datasets.

---
[‡] https://www.enexis.nl/over-ons/open-data

### 3.2.1 Overall performance and capacity classification

Figure 9 classifies each neighborhood into one of three capacity ranges – below 1000 kW$_p$ (blue points), between 1000 to 2000 kW$_p$ (orange points), and above 2000 kW$_p$ (red points) – allowing us to assess how well the model scales from relatively low to high PV penetration areas. As shown, most of points lie near or on the diagonal, indicating strong overall agreement with the DSO records. The linear regression fit yields an $R^2$ value of 0.92, indicating that ∼92% of the variance in the recorded data is captured by our predictions. In addition, the Mean Absolute Error (MAE) across all neighborhoods stands at 138.09 kW$_p$, while the Mean Absolute Percentage Error (MAPE) is 26.59%. These metrics suggest that the method robustly reproduces neighborhood-level capacity pattern as documented by the DSO, especially considering the complexity of rooftop-level detection over large urban regions.

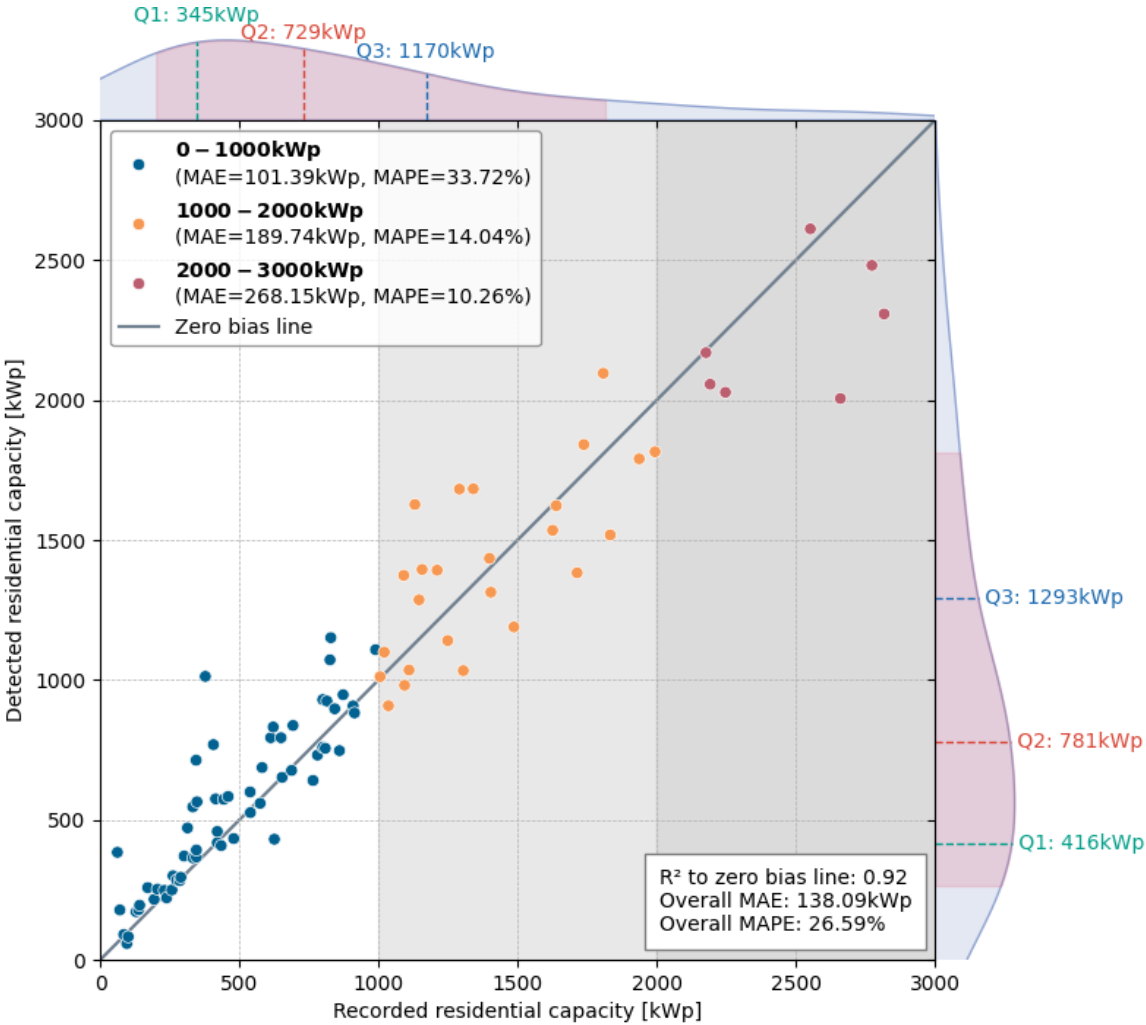

Figure 9. Scatter statistics of neighborhood-level residential PV capacity estimation versus ground truth recordings with acceptable bias margins.

Exploring the recorded capacity distribution further, the three quartiles of recorded data appear at 345 kW$_p$, 729 kW$_p$, and 1170 kW$_p$, while the corresponding model quartiles are 416 kW$_p$, 781 kW$_p$, and 1293 kW$_p$. The detection percentage error at these quartiles remains relatively moderate, range from around 7% to 20%, suggesting a stable performance across the interquartile range, regardless of whether a neighborhood has a modest or substantial number of PV installations.

Furthermore, dividing the data by capacity range reveals that, for neighborhoods above 2000 kW$_p$, the model's absolute error grows (MAE ≈ 268.15 kW$_p$) but the relative error remains small (MAPE≈10.26%), confirming that the approach appropriately scales up to higher installation volumes. For neighborhoods in the low-capacity bracket (< 1000 kW$_p$), the MAE decreases (101.39 kW$_p$) but the MAPE increases (33.72%) because the denominator (the true installed capacity) is smaller. These findings collectively indicate that while absolute errors can inflate in neighborhoods containing more extensive PV capacity, the fraction of deviation stays modest. Conversely, in neighborhoods with lower rooftop installations, the MAPE metric can appear elevated despite small real differences because the total installed capacity itself is limited. Overall, Figure 9 demonstrates that the model performs reliably across a diverse spectrum of PV penetration levels when compared against the available DSO data.

### 3.2.2 Acceptable bias margins and error distributions

Figure 10 introduces two dashed lines that delineate a ±25% margin relative to the diagonal. Any point falling between these lines exhibits an Absolute Percentage Error (APE) within 25%. The selection of this 25% threshold is informed by two key considerations. Firstly, as discussed in Section 2.2.3, the inherent accuracy limitations of mask-to-polygon conversion (IoU at 0.3m GSD around 0.76) can naturally lead to discrepancies in PV area detection, suggesting that capacity estimates might reasonably diverge by a similar fraction. Secondly, and critically, this margin aligns remarkably well with findings from external research. For instance, a Dutch investigation found a country-wide discrepancy of approximately 25% between registered and actually installed PV capacity [81]. This congruence suggests that a ±25% margin is not only reflective of our model's expected performance envelope but also accounts for a realistic level of uncertainty and potential under- or over-registration within the reference DSO data itself. Therefore, this margin helps define a zone of "plausible agreement" rather than strict accuracy against a potentially imperfect truth.

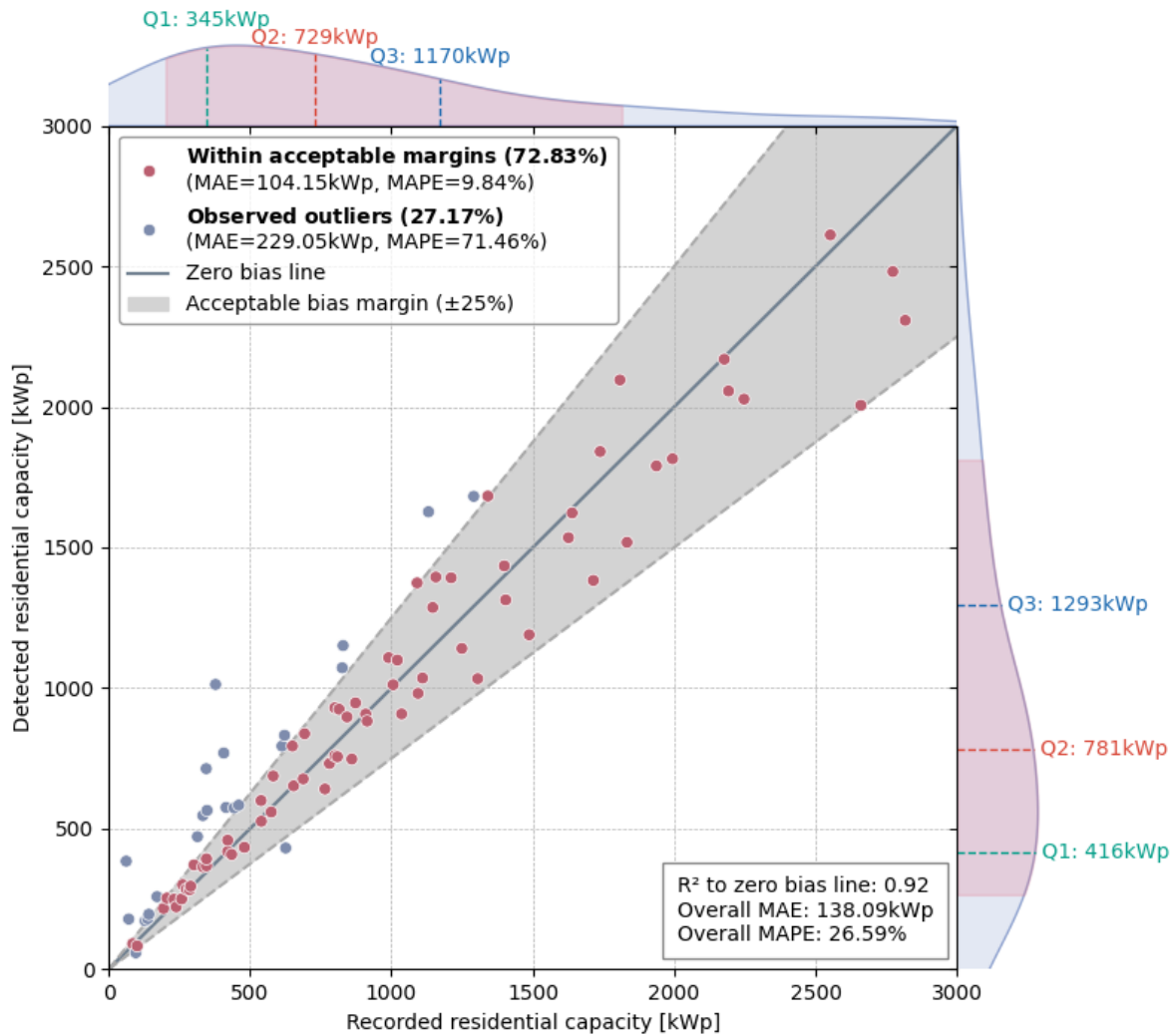

Figure 10. Scatter statistics of neighborhood-level residential PV capacity estimation versus ground truth recordings with acceptable bias margins.

Results confirm this expectation: approximately 72.83% of Eindhoven's neighborhoods fall inside the ±25 % zone. For these "in-range" points, which can now be interpreted with greater confidence as accurately estimated neighborhoods given the context of reference DSO data uncertainty, the MAE and MAPE are 104.15 $kW_p$ and 9.84 %, respectively, attesting to strong consistency with the DSO's records. For the remaining 27.17% of neighborhoods, the model over- or underestimates by a more substantial margin (MAE≈229.05 $kW_p$, MAPE≈71.46%). A deeper investigation reveals several potential causes of these larger discrepancies. First, as inherent to remote sensing analysis, darker patches – whether caused by water pooling, shadows, or certain roof materials – can mimic PV modules in overhead imagery, leading to inflated capacity estimates. Conversely, real modules installed on similarly dark-colored rooftops can remain underdetected if their spectral signatures blend into the background. Given that single-timestamp satellite imagery cannot readily avoid transient weather conditions or solar beam-induced shadows, some over- or underestimation is unavoidable without more advanced corrections or multispectral data. Furthermore, it is also plausible that some of these larger deviations could, in part, reflect more significant local inaccuracies or outdated information within the DSO's registration data for those specific neighborhoods. Nonetheless, the majority of neighborhoods remain unaffected by these complexities, suggesting that the core parameterization pipeline is fundamentally robust.

In summary, Figures 9 and 10 confirm that the proposed automated PV parameterization framework excels at retrieving neighborhood capacity patterns across the broad urban area. The estimates are not only largely consistent with official DSO records but, significantly, the majority fall within a well-justified margin of agreement that acknowledges the inherent uncertainties in both the remote sensing-based estimation and the reference data itself. This positions the framework as a valuable tool for providing an independent, up-to-date assessment of urban PV deployment, which can complement and potentially help identify areas for further verification in official DSO registries, thereby aiding decisions on urban PV deployment and grid management.

## 4. Application in system production profile modeling

The primary aim of our automated urban PV parameterization framework is not only to locate and characterize urban PV systems but also to support high-fidelity power-generation simulations – a crucial aspect for distribution grid operators, energy planners, and other stakeholders who require accurate, data-driven insights into PV production. In this section, we demonstrate how the detailed parameters generated by our proposed framework seamlessly integrate with two established PV performance models, highlighting how advanced simulations become far more precise and insightful when array-specific details, derived from our novel estimation techniques are available.

### *4.1 Demonstration scene*

The demonstration focuses on a local community in Eindhoven, comprising 38 residential dwellings within a single low-voltage distribution loop. As detailed by Figure 11, of these 38 buildings, 27 contain diverse-oriented rooftop PV arrays, resulting in a relatively high penetration ratio for a residential neighborhood. This high PV penetration renders the local grid particularly sensitive to short-term variations in solar output, as significant voltage fluctuations and reverse power flows can readily arise during periods of strong sunshine. Such an environment, with its inherent variability in PV system configurations (tilt, azimuth, and potential inter-building/vegetation shading) – compounded by these grid sensitivities – represents a typical complex urban setting where our detailed parameterization framework offers significant advantages. From a grid operator's standpoint, oversimplifying these PV configurations – by assuming uniform tilt, orientation, or capacity, a common necessity when detailed data is absent – can lead to substantial forecasting errors in local generation and may mask the true potential for congestion or voltage management challenges.

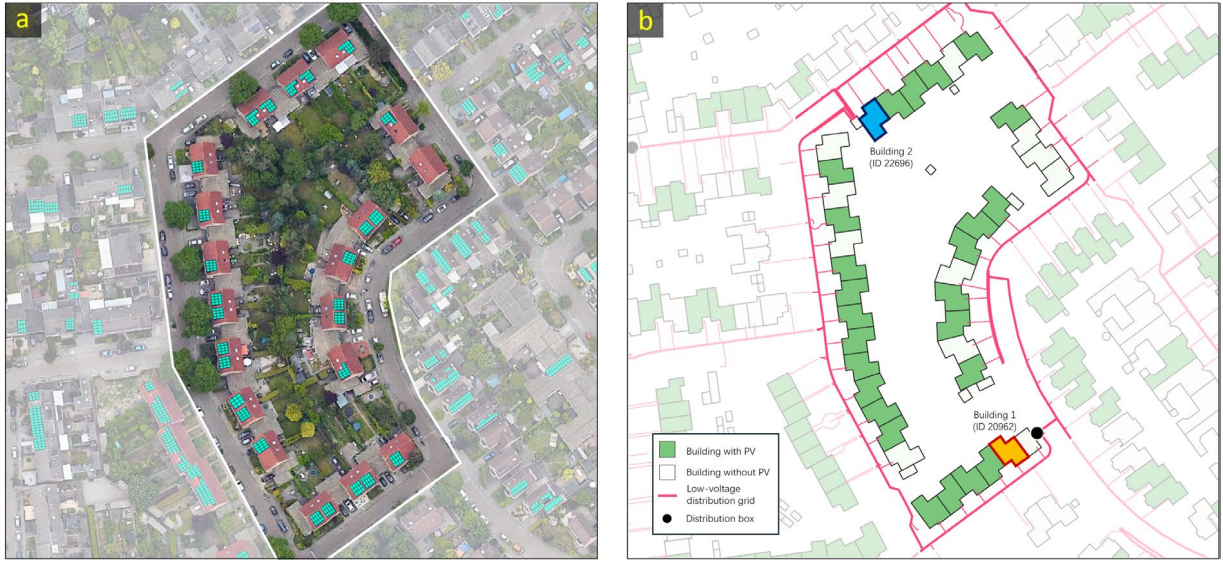

Figure 11. Overview of the demonstration community, wherein (a) satellite imagery with detected PV array positions and module layouts; (b) Translation of the detected PV installation information into building-level GIS layer, the bounded low-voltage distribution grid loop is highlighted.

To evaluate the importance of including detailed installation data captured by our framework in real-world system modeling, we select a one-week period in April 2022 characterized by predominantly clear-sky weather, when high irradiance can induce significant midday power export to the grid. The short-term variations and potential partial-shading effects in the dense urban setting make an especially compelling case for deploying advanced, configuration-aware simulations powered by the granular outputs of our framework. Meteorological inputs for this time window were retrieved from the Koninklijk Nederlands Meteorologisch Instituut (KNMI) [82] and include solar irradiance, ambient temperature, and wind speed.

### *4.2 PV parameterization – performance model connection*

Because the proposed framework captures not only the presence of PV but also the specific geometric and module-level configuration of each installation, a seamless connection to detailed power models is possible. For each building, the parameters derived by our framework – recognized tilt, azimuth, rated module specifications, and spatial layout – are transferred directly into two widely adopted simulation packages:

- *PVlib*. As one of the most widely used toolkits in academic and industry settings [83], PVlib offers the Sandia Array Performance Model (SAPM) for simulating DC output under varying irradiance and temperature. Parameters such as tilt, azimuth, and module properties extracted by our framework directly feed into SAPM. However, it does not explicitly account for complex shading effects or mismatch among modules. Hence, its predictions often represent a potential optimistic scenario of PV output.

- *PVMismatch (PVMM)*. A more advanced framework that models the current-voltage (IV) characteristics of each module and includes provisions for simulating mismatch across a string [84]. Because the explicit string arrangement is not accessible from overhead imagery, PVMM treats all modules in a given PV array as connected in a single string, creating a potential pessimistic scenario for mismatch losses if a portion of the array becomes shaded. The detailed module layout provided by our framework is crucial for such fine-grained modeling.

  Taken together, these two models, when fed with the detailed parameters from our framework, produce a range of possible generation profiles – an upper bound (PVlib) where shading is largely ignored, and a lower bound (PVMM) where the entire array is susceptible to mismatch from localized shading effects. In real-life operations, the actual energy yield would likely fall somewhere between these scenarios (bounds), depending on the precise wiring configuration and local environment.

Additionally, to highlight the gains from using detailed parameterization data, we also generate two "baseline" energy production profiles. These Baselines reflect commonly used assumptions in the state-of-the-art when such detailed, array-specific data are unavailable, forcing reliance on generalizations [85]:

- *Baseline 1*. All modules are assumed to face due south (azimuth=180°) at a fixed optimal tilt (local latitude minus 15°, for the Netherlands is roughly 35°), without shading effects.
- *Baseline 2*. All modules are assumed flat (tilt = 0°), without shading effects.

These default configurations neglect potential orientation, tilt, and shading diversity, thereby introducing considerable uncertainties in any resulting generation estimates. In prior large-scale studies, such approximations were often deemed inevitable – data regarding rooftop PV specifics were simply unavailable at scale. By contrasting the high-fidelity outputs generated via our workflow with these baseline results, we illustrate both the limitations of legacy assumptions and the substantial accuracy gains offered by our automated parameterization approach.

### *4.3 Building-level production profile*

We begin by diving into building-level production profiles, from which we can explore how roof orientation, module layout, and shading nuances affect hourly PV output – insights that sometimes become obscured in aggregated analyses. Two sample buildings are selected (details in Figure 11b) to illustrate contrasting shading conditions: Building 1 has minimal tree or structural interference (Figure 12a), while Building 2 faces a dense tree cluster (Figure 12b). This contrast helps demonstrate how partial shading expands or narrows the gap between optimistic (non-shading) and more realistic (shading-aware) performance forecasts, thereby emphasizing the importance of configuration-specific modeling enabled by our detailed parameterization.

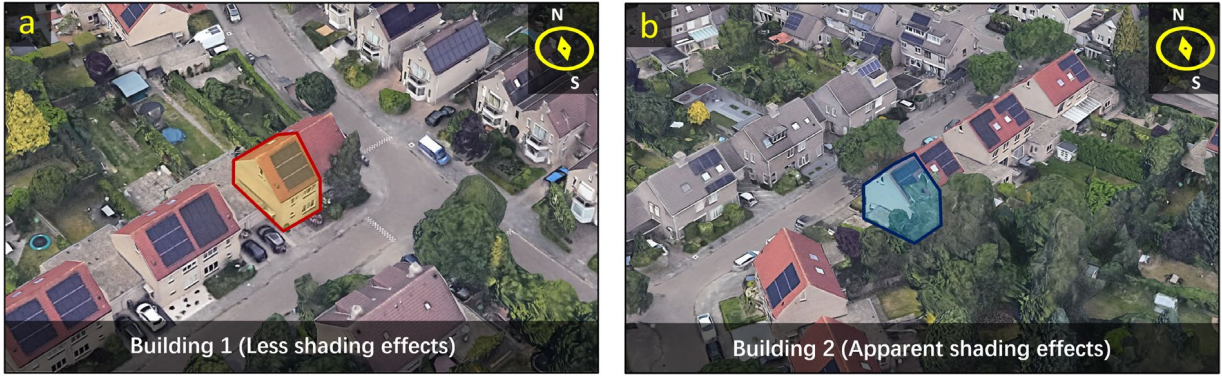

Figure 12. Visualizations of the two selected buildings and their surrounding settings, wherein (a) Building 1 and (b) Building 2.

Figures 13 and 14 illustrate the time-based power output from the two sampled buildings. In each figure, two main simulation profiles – those from PVlib (upper bound) and PVMM (lower bound) – form a Generation Predictive Band (GPB) that highlights how hourly energy output may vary when shading and electrical mismatch are either ignored or accounted for. The width of GPB at each timestamp $i$ indicates the possible deviation of realistic energy production from the optimistic non-shading situation. Two additional dashed curves represent "baseline" assumptions that omit all site-specific details, thereby illustrating how large-scale modeling often errs when tilt, azimuth, or partial shading are unknown.

To systematically compare model outputs, we employ several quantitative measures over entire time window of $N$ timestamps. Let $P_{pvlib}(i)$ denote the power predicted by PVlib, and $P_{pvmm}(i)$ the power from PVMM, we define the Mean Absolute Percentage Width (MAPW) and Cumulative Percentage Width (CPW) for an established GPB as

$$MAPW = \frac{100\%}{N} \sum_{i=1}^{N} \frac{|P_{pvlib}(i) - P_{pvmm}(i)|}{P_{pvlib}(i)} \qquad (4)$$

$$CPW = \frac{\sum_{i=1}^{N}(P_{pvlib}(i) - P_{pvmm}(i))}{\sum_{i=1}^{N} P_{pvlib}(i)} \times 100\% \qquad (5)$$

MAPW intends to show the degree of uniformed power reduction at the individual timestamp, which is crucial for quantifying shading effects in short-term production modeling tasks. Conversely, CPW shows the total energy loss as a percentage of maximal production potential within a given time window, which is more relevant for assessing shading effects in long-term modeling practices.

To evaluate how closely a baseline scenario $B_{base}(i)$ tracks the upper and lower bounds of GPB, we calculated the Mean Absolute Percentage Error with respect to PVlib ($MAPE_H$) and PVMM ($MAPE_L$) as

$$MAPE_H = \frac{100\%}{N} \sum_{i=1}^{N} \frac{|B_{base}(i) - P_{pvlib}(i)|}{P_{pvlib}(i)} \qquad (6)$$

$$MAPE_L = \frac{100\%}{N} \sum_{i=1}^{N} \frac{|B_{base}(i) - P_{pvmm}(i)|}{P_{pvmm}(i)} \qquad (7)$$

Subsequently, we define the Cumulative Percentage Error of baseline scenarios with respect to PVlib ($CPE_H$) and PVMM ($CPE_L$) as

$$CPE_H = \frac{\sum_{i=1}^{N}(B_{base}(i) - P_{pvlib}(i))}{\sum_{i=1}^{N} P_{pvlib}(i)} \times 100\% \qquad (8)$$

$$CPE_L = \frac{\sum_{i=1}^{N}(B_{base}(i) - P_{pvmm}(i))}{\sum_{i=1}^{N} P_{pvmm}(i)} \times 100\% \qquad (9)$$

Analogous to the MAPW and CPW that reveal the range between optimistic and pessimistic performance estimates (introduced by GPB), MAPE and CPE illustrate how a simplified baseline assumption deviates from the predictions made with site-specific parameters from our framework. Nonetheless, MAPE uncovers the magnitude of error fluctuations for each individual timestamp, while CPE summarizes errors over the entire timeframe.

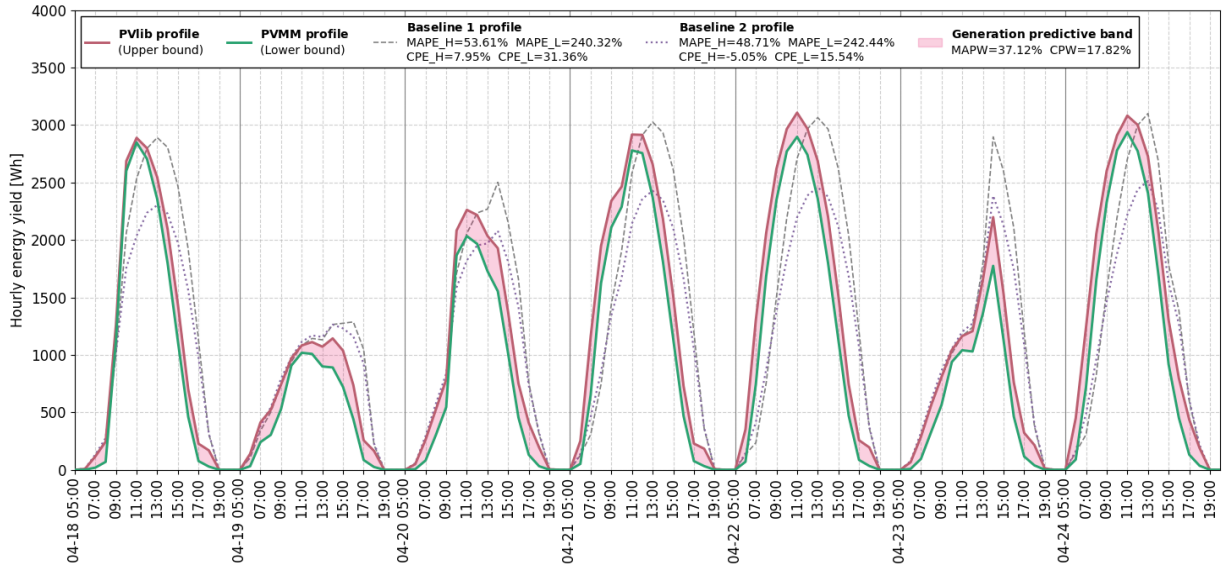

Figure 13. Modeled hourly PV production profile of Building 1 with minor shading effects.

Figure 13 details the daily power cycles for Building 1, which experiences minimal external shading. The PVlib (red curve) and PVMM (green curve) outputs remain close during most daylight hours, forming a comparatively narrow GPB. On the days with clear sky (near-peak irradiance), there is often a consistent alignment between the two curves. However, in situations of mostly diffuse light – such as the cloudy midday window on April 19 and 20 – PVMM values are noticeably below PVlib, at times by up to 250 Wh. This observation aligns with the findings from prior comparative studies [86–88], reflecting differences in how each model handles low-irradiance conditions. The two-diode model adopted by PVMM, being more physically detailed to account for mechanisms like shunt resistance losses that are prominent during shading or weak sunlight, yields lower predicted power compared to the empirical models (e.g., SAPM model) by PVlib. Overall, considering both the converged results in high-irradiance period and the diverse results in low-irradiance periods, Building 1 exhibits a MAPW of about 37.12% and a CPW of 17.82%, suggesting that, despite mild shading, there remains a nontrivial difference between fully idealized predictions and those accounting for subtle mismatch and explicit circuit behaviors.

Comparisons with the two baseline curves in Figure 13 highlight how omitting site-specific installation data produces temporal shifts and magnitude errors in the predicted power. Baseline 1 assumes a south-facing tilt of 35°, causing its daily power peaks to occur roughly two hours later than the actual (southeast) orientation. Although Baseline 1's peak magnitude can occasionally resemble the GPB's upper boundary (around 3000 Wh on sunny afternoons), the delayed timing distorts any short-term forecasts, leading to a $MAPE_H$ at 53.61% and a $MAPE_L$ at 240.32%. Even focusing on the long-term forecast scope, the mismatch remains evident, with Cumulative Percentage Errors of 7.95% and 31.36% for PVlib ($CPE_H$) and PVMM ($CPE_L$), respectively.

Meanwhile, Baseline 2 (flat orientation) fails to approach the GPB's peak outputs, reflecting its zero-tilt assumption. Although its diurnal pattern can merge briefly with the GPB on heavily overcast mornings (when orientation matters less), Baseline 2 systematically underestimates midday peaks, incurring $MAPE_H$=48.71% and $MAPE_L$=242.44%. However, because periodical under- and over-estimations are mutually offset, Baseline 2 presents smaller $CPE_H$ (at -5.05%) and $CPE_L$ (at 15.54%) than Baseline 1, implying an improved long-term prediction performance. Overall, these baseline results confirm that, even under modest shading, failing to capture actual system configuration data can cause large forecasting errors that are both temporal (shifted peak times) and quantitative (inaccurate power magnitudes).

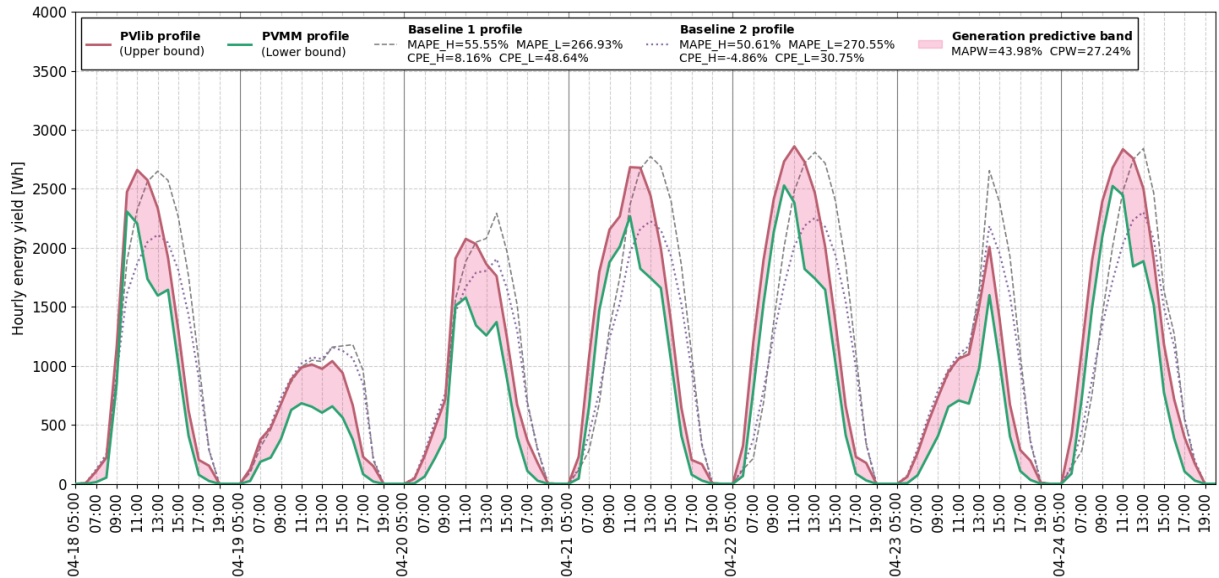

Figure 14. Modeled hourly PV production profile of Building 2 with substantial shading effects.

Unlike Building 1, Building 2 sits adjacent to multiple mature trees, creating substantial partial shading. Figure 14 reveals a broader GPB, with a MAPW of 43.98% and an CPW of 27.24%. On clear-sky days, shading becomes most pronounced between roughly 10:00 and 14:00, where PVMM can register nearly 900 Wh (about 30 %) less output than PVlib's non-shading assumption. Even during cloudier periods, the gap typically remains around 500 Wh at midday, further illustrating how dynamic shading losses accumulate over a range of irradiance conditions. This divergence reflects the practical necessity of high-resolution data on array settings and environmental obstructions: while ignoring shading leads to an overly optimistic power forecast, the precise detection and array layout information from our framework enables advanced models like PVMM to quantify mismatch losses that can sometimes dominate performance in tree-lined neighborhoods.

As with Building 1, the two baseline scenarios of Building 2 in Figure 14 show more pronounced discrepancies in both shape and timing when compared to the GPB. Baseline 1, despite matching the GPB's upper envelope under strong solar irradiance, again lags by around two hours in peak production. Its MAPE is increased to 55.55 % relative to PVlib and 266.93 % relative to PVMM, magnifying how orientation errors alone can drastically shift model outcomes, especially in the face of shading. Baseline 2 stands even further from the GPB, demonstrating a lower-peaked diurnal curve that departs from either advanced model's predictions. In addition, compared to the shading-free case (Figure 13), the cumulative error of Baseline 2 is also notably enlarged, especially with respect to the PVMM results ($CPE_L$), which nearly doubles to 30.75%. Ultimately, Building 2 shows that partial shading introduces additional error drivers on top of orientation or tilt differences – system performance modeling becomes not merely a matter of capturing correct angles but also responding to local obstructions that degrade output. Under such conditions, conventional large-scale assumptions can prove dangerously misleading in operational or planning contexts.

Across these two building-level demonstrations, the gap between non-shading and shading-aware models (i.e., the width of GPB) grows in proportion to nearby obstruction complexities. The errors introduced by baseline assumptions reveal that neglecting array-specific tilt, azimuth, and layout – key parameters delivered by our framework – can lead to apparent temporal shifts and systematic over- or underestimations of actual output. These observations reinforce the indispensability of our automated workflow, which enables advanced power simulations like PVMM to be deployed at scale and provides improved energy forecasting accuracy even under varying shading scenarios.

### 4.4 Community-level production profile

Having explored the variability of single-building simulations, we now broaden our view to the collective PV output of the entire demonstration community within the distribution loop (see Figure 11 b). This shift in focus allows us to investigate whether the shading and orientation effects evident at smaller scales also persist at the aggregate level – and, more critically, how they might affect real-world applications such as distribution grid planning and operational decision-making, when informed by detailed, system-specific data from our framework.

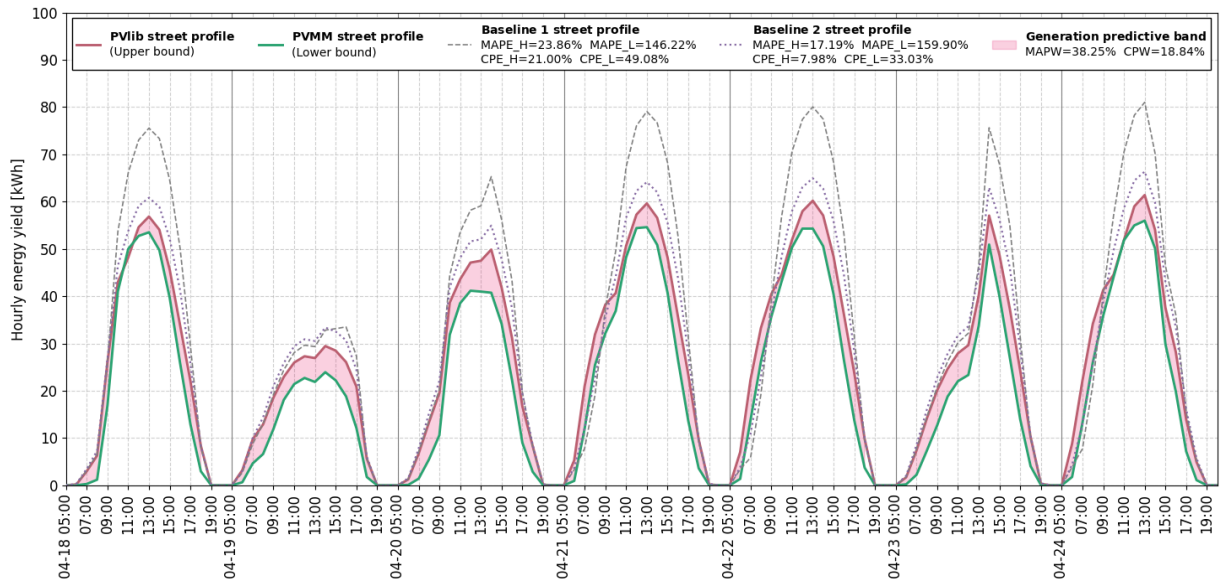

Figure 15. Modeled hourly PV production profile of the entire demonstration community.

A key question for community-level modeling is whether the specificity of individual rooftop systems (e.g., partial shading from trees or dormers) substantially affects the total output once the profiles are summed. As Figure 15 shows, the power trajectories from PVlib and PVMM continue to diverge in a manner similar to single-building cases, albeit with somewhat muted extremes. Under sunny midday conditions, the shading-aware PVMM output can run about 10% below the non-shading PVlib estimate; under cloudier, more diffuse conditions, the difference may rise to around 20%. These discrepancies create a GPB whose width is in between the previous two individual building cases, with moderate MAPW at 38.25%. In practice, this means that ignoring system-specific shading factors leads to an over-optimistic forecast of the entire community's time-stamped production – even when unshaded buildings are present to offset part of the deficit.

Moreover, the timing of shading losses becomes more continuous at the aggregated level. In single-building analyses (e.g., Building 2), the most pronounced gaps between PVlib and PVMM occurred within a specific window (roughly 10:00 to 14:00). At the community level, however, systems with different orientations or upfront obstructions experience partial shading at different hours. Consequently, gaps between the two GPB's bounds periodically widen and narrow over the daylight hours, reflecting how the sum of many staggered, smaller shading events yields a persistent mismatch. As highlighted by CPW, the continuous daylight time shading ultimately cumulated to a notable energy loss at 18.84%. From an application standpoint, this finding reveals that local shading effects do not merely "average out" across numerous systems; distribution operators evaluating midday loading, reverse power flows, and voltage regulation must still account for shading-induced output deficits in aggregated scenarios.

As for the two baseline assumption scenarios, Figure 15 confirms that, once again, these assumptions deviate considerably from either the upper (PVlib) or lower (PVMM) bounds generated with actual system configurations. Interestingly, the community-level baseline profiles now show midday peaks that align more closely in time with the GPB's real aggregate peak, in contrast to the building-level results where a uniform orientation consistently lagged or led by about two hours. This difference is due to the artifact of diverse building orientations: when some rooftops face southeast and others face southwest, their individual timing errors can partially cancel out in total. The mitigated timing errors further reduced the mismatch between baseline assumptions and GBP envelopes – for Baseline 1, we observe MAPE of 23.86 % and 146.22 % relative to PVlib and PVMM, respectively. Baseline 2 fares similarly, with $MAPE_H$=17.19 % and $MAPE_L$=159.90 %.

Crucially, these baselines still overestimate the total output, surpassing the GPB peak by a significant margin. Because baseline profiles stack up their production at one common midday maximum, the aggregated peak is inflated relative to the more staggered (and partially shaded) system-specific profiles. Consequently, both baseline profiles exhibit enlarged cumulative error on the community scale. Compared to building-level results in Figure 13, the $CPE_H$ and $CPE_L$ of Baseline 1 increased by over 15% to 21.00% and 49.08%, respectively. The change of Baseline 2 is similar to that of Baseline 1, with the increase of $CPE_H$ to 7.98% and $CPE_L$ to 33.03%. In other words, while orientation errors can occasionally offset each other in terms of timing, baseline assumptions still inflate total peak production capacity due to a false synchronization of individual building outputs – leading to noticeable performance discrepancies that hamper both short- and long-term energy production forecasting.

In summary, the community-level results reiterate the critical role of configuration-aware modeling, which is directly enabled by our automated PV parameterization framework. Although the shading losses from individual buildings cannot fully dominate the regional aggregated outputs, localized obstructions remain sufficient to produce a persistent mismatch between optimistic and realistic forecasts. At the same time, default baselines can notably overestimate the energy production – a misleading outcome that can distort grid expansion planning or operational strategies. Our framework provides the necessary detailed inputs to move beyond such problematic generalizations.

## 5. Discussion

### 5.1 Significance

The end-to-end urban PV parameterization framework presented in this work – encompassing system segmentation, innovative array footprint vectorization, tilt-azimuth estimation, and detailed module layout inference – marks a critical step toward more robust urban PV data ecosystems. Drawing on the empirical results from our experiments, we highlight two key dimensions of significance.

#### 5.1.1    Enriched urban-scale PV datasets for diverse energy applications

The central strength of this framework lies in its ability to generate an all-in-one GIS layer detailing the precise footprint, tilt, azimuth, and layout of each rooftop array across an entire urban area. As demonstrated in our building-level (Sections 4.3) and community-level (Section 4.4) analyses, such granular data is pivotal for moving beyond simplified modeling assumptions and capturing the full range of real-world variability – particularly when shading issues drive significant disparities between optimistic (non-shading) and more realistic (shading-aware) forecasts.

Equipped with accurate GIS-based PV data, distribution system operators can better anticipate load flows, voltage constraints, and potential congestion during periods of peak solar output. Our case study demonstrated that ignoring or oversimplifying system-specific details can lead to inflated midday peaks or shifted production curves, which jeopardizes both operational planning and investment decisions. By contrast, an automated, data-driven inventory of all rooftop systems, as provided by our parameterization framework, empowers grid operators to identify the realistic range of possible production patterns (e.g., the GPB discussed in Sections 4.3-4.4), thus invest in targeted reinforcement or adopt dynamic control schemes (e.g., demand-response, energy storage) precisely where they are most needed.

The utility of this enriched dataset also extends well beyond grid management. Urban planners can integrate the roof-specific solar metrics into broader sustainability targets, track progress toward net-zero goals. Likewise, municipal decision-makers can better identify rooftops ripe for additional solar retrofits or prioritize incentives for neighborhoods with suboptimal configurations. Moreover, advanced simulation tools – such as PVMismatch – become feasible at scale, translating our framework's detailed outputs directly into shading-aware circuit models rather than lump-sum capacity estimates. In essence, by bridging raw imagery with in-depth PV system parameterization, the proposed framework provides the granular data backbone for virtually any domain where rooftop solar matters.

### 5.1.2 Scalable and transferable workflow

A second major significance is the scalable design of the proposed framework. Building on widely available data sources – high-resolution satellite images and elevation (LiDAR) data – the modular pipeline is executable in a full-automated manner. By demonstrating robust segmentation performance at multiple ground sample distances (Section 2.2) and successfully applying the same pipeline logic in both urban-level and community-level (Sections 3-4), we confirm that the approach extends beyond a single scale or image source.

Moreover, the transferability goes beyond replicating methods in different localities. The system's adaptability – e.g., straightforward reconfiguration of segmentation networks, uniform data-structuring of polygon outputs, and flexible module-library matching – paves the way for cross-country or multiregional projects. This is especially relevant where rapid PV expansion necessitates up-to-date, detailed registries of distributed generation.

By uniting these two pillars – finer-grained PV data derived from advanced parameterization tailored to stakeholder needs and an architecture that scales with minimal retraining or local customization – the proposed workflow offers a robust means to continuously maintain and leverage rich rooftop PV databases across diverse urban contexts.

### *5.2 Limitations and outlooks*

Although the proposed framework demonstrates considerable promise for mapping and modeling rooftop PV, the validation study and subsequent analyses also reveal inherent constraints and opportunities for further refinement. In the demonstration study for Eindhoven (Section 3), for instance, around 73 % of neighborhoods matched the recorded DSO data within $\pm 25$ % – a robust indication that the method is well suited for city-level analyses. At the same time, the remaining 27 % of cases highlight circumstances where transient conditions or atypical rooftop materials generate false positives or underdetections. Water puddles, shadows cast by nearby structures, or dark roofing that closely resemble PV surfaces can confound the segmentation network's predictions. To address these misclassifications, future research could integrate multi-temporal satellite imagery (thereby reducing the influence of short-term artifacts) or leverage color-band decomposition (i.e., aggregate color features tailored to PV materials) to improve the module detection from visually similar roof materials.

Another limitation pertains to the workflow's current focus on rooftop-mounted PV arrays. As solar adoption grows more diverse, vertically installed systems or façade-integrated modules will become increasingly relevant in high-density environments. Our methodology, which relies on overhead imagery, is not yet equipped to detect these vertical arrays, thus excluding a potentially significant share of future deployments in urban areas. One potential enhancement is to incorporate stereo-photogrammetry or 3D reconstruction data (e.g., from LiDAR scans or street-view imagery) so that façades and other vertically-tilted surfaces can be more comprehensively mapped and included in citywide solar assessments.

A third consideration arises when bridging the detailed PV data layer from our framework to advanced performance models such as PVMismatch. Although modeling each rooftop individually provides insights into shading and mismatch, the computational cost grows quickly at large scales. For instance, at the community-scale (as we presented in Section 4), with multiple rooftop arrays sharing similar orientations and minimal shading, repeatedly simulating each system's detailed electrical characteristics may be redundant. Grouping arrays with comparable configurations and modeling them as "equivalent arrays" could streamline large-scale production forecasts, reducing computational overhead without sacrificing accuracy. Developing and validating such an equivalence-based approach is a natural avenue for future research, particularly where timely simulations are needed for real-time grid operations.

Overall, continued improvements along these lines will strengthen the parameterization pipeline's adaptability to shifting diverse urban landscapes and future developments, ensuring that the resulting PV data layers remain robust, comprehensive, and directly relevant to the operational and planning challenges faced by modern cities.

# 6. Conclusion

This study introduced and validated a fully automated framework for detecting and retrieving rooftop PV installation details in urban settings. By combining high-resolution remote sensing data, deep-learning-based segmentation, polygon-vector conversion, elevation-based tilt-azimuth estimation, and detailed module layout inference, we produced a richly attributed dataset of distributed PV systems. In validation tests for Eindhoven (the Netherlands), approximately 73% of neighborhoods registered capacity estimates within $\pm 25\%$ of recorded DSO data, and the correlation ($R^2$) between the model's outputs and operator data reached 0.92 – indicating that this parameterization pipeline effectively bridges the data gap that long hindered large-scale, site-specific analysis of urban PV installations.

A defining strength of this framework is its ability to integrate seamlessly with detailed performance models such as PVlib and PVMismatch. Rather than resorting to broad assumptions (e.g., uniform 0° or 35° tilt), each array in the dataset carries automatically derived parameters – including orientation, tilt, and module layout – allowing simulations to capture high-fidelity system behaviors. The resulting Generation Predictive Band (GPB) delineates realistic power production ranges under various working conditions. At the single-building scale, compared to GPB, modeling PV performance with conventional simplified assumptions (which lack these detailed parameters) often wrongly estimated the time-stamped power outputs and introduced substantial timing errors (e.g., two-hour offsets in daily peaks). Ultimately, due to these omitted details, the Mean Absolute Percentage Errors for short-term forecasts went up to 270% and Cumulative Percentage Error for long-term forecasts up to 48%. Even at the community level, summing many diverse rooftops did not nullify these significant discrepancies, with baseline assumption profiles cumulatively biased from the realistic configuration-aware profiles up to 50 %. Thus, by replacing simplified assumptions with detailed system-specific data, the proposed method fundamentally progresses the large-scale PV modeling practices.

Looking forward, the modular structure and adaptability of this workflow make it broadly transferable across regions and scales. As growing PV penetration prompts increasing interest in accurate solar forecasts – from day-ahead scheduling to long-term infrastructure planning – the ability to repeatedly update detailed rooftop PV inventories with minimal human intervention becomes invaluable. Armed with the comprehensive PV GIS layers generated here, stakeholders can more confidently track expansion trends, refine grid management strategies, and align policy measures with actual rooftop potential.

# Acknowledgments

The present work is sponsored by the scholarship from China Scholarship Council (CSC, 202107720038).

# Appendix A

## *A.1 List of standard CEC approved module templates*

| Index [-] | Label [-] | Height [mm] | Width [mm] | Cell number [-] | Material [-] |
|---|---|---|---|---|---|
| 0 | Mono-c-Si_0.017_128 | 2067 | 1046 | 128 | Mono-c-Si |
| 1 | Mono-c-Si_0.017_96 | 1559 | 1046 | 96 | Mono-c-Si |
| 2 | Mono-c-Si_0.022_72 | 1620 | 980 | 72 | Mono-c-Si |
| 3 | Mono-c-Si_0.025_72 | 1924 | 954 | 72 | Mono-c-Si |
| 4 | Mono-c-Si_0.026_60 | 1620 | 980 | 60 | Mono-c-Si |
| 5 | Mono-c-Si_0.026_72 | 1893 | 971 | 72 | Mono-c-Si |
| 6 | Mono-c-Si_0.026_96 | 1943 | 1297 | 96 | Mono-c-Si |
| 7 | Mono-c-Si_0.027_60 | 1644 | 979 | 60 | Mono-c-Si |
| 8 | Mono-c-Si_0.027_72 | 1966 | 1000 | 72 | Mono-c-Si |
| 9 | Mono-c-Si_0.027_96 | 1980 | 1300 | 96 | Mono-c-Si |
| 10 | Mono-c-Si_0.028_60 | 1680 | 996 | 60 | Mono-c-Si |
| 11 | Mono-c-Si_0.028_72 | 2000 | 1000 | 72 | Mono-c-Si |
| 12 | Mono-c-Si_0.033_60 | 1960 | 998 | 72 | Mono-c-Si |
| 13 | Multi-c-Si_0.024_72 | 1864 | 932 | 72 | Multi-c-Si |
| 14 | Multi-c-Si_0.025_72 | 1924 | 954 | 72 | Multi-c-Si |
| 15 | Multi-c-Si_0.026_60 | 1614 | 954 | 60 | Multi-c-Si |
| 16 | Multi-c-Si_0.026_72 | 1934 | 970 | 72 | Multi-c-Si |
| 17 | Multi-c-Si_0.026_96 | 1943 | 1297 | 96 | Multi-c-Si |
| 18 | Multi-c-Si_0.027_60 | 1640 | 998 | 60 | Multi-c-Si |
| 19 | Multi-c-Si_0.027_72 | 1970 | 988 | 72 | Multi-c-Si |
| 20 | Multi-c-Si_0.028_60 | 1670 | 1000 | 60 | Multi-c-Si |
| 21 | Multi-c-Si_0.028_72 | 1994 | 1000 | 72 | Multi-c-Si |
| 22 | Multi-c-Si_0.033_60 | 1960 | 998 | 60 | Multi-c-Si |